\documentclass[
  floats,
  floatfix,
  reprint,
  superscriptaddress,
  nofootinbib,
  longbibliography,
]{revtex4-2}
\usepackage{amsfonts,amssymb,amsmath}
\usepackage[caption=false]{subfig} %
\usepackage{graphicx, hyperref, tabularx, dcolumn}

\usepackage{color}
\usepackage[dvipsnames]{xcolor}
\usepackage{soul}
\usepackage[normalem]{ulem}
\usepackage{orcidlink}
\usepackage{tensor, comment}
\usepackage{enumitem}
\usepackage{multirow}
\usepackage{CJKutf8}

\newcommand{\dual}{\,{}^*\!}

\newcommand{\e}[1]{\times 10^{#1}}

\newcommand{\athenak}{\texttt{AthenaK}}

\defcitealias{Most:2024onq}{MW25}

\definecolor{ykcolor}{rgb}{0.8, 0, 1}

\graphicspath{{./}{Figures/}}

\newcommand{\caltech}{\affiliation{Theoretical Astrophysics 350-17, California
Institute of Technology, Pasadena, CA 91125, USA}}
\newcommand{\burke}{\affiliation{Walter Burke Institute for Theoretical Physics,
California Institute of Technology, Pasadena, CA 91125, USA}}

\begin{document}
\begin{CJK*}{UTF8}{gbsn}

\author{Yoonsoo Kim \orcidlink{0000-0002-4305-6026}} 
\email{yoonsoo@princeton.edu}
\affiliation{Princeton Center for Theoretical Science, Princeton University, Princeton, NJ 08544, USA}
\affiliation{Princeton Gravity Initiative, Princeton University, Princeton, NJ 08544, USA}
\caltech

\author{Elias R. Most \orcidlink{0000-0002-0491-1210}}
\email{emost@caltech.edu}
\caltech
\burke

\author{Hai-Yang Wang (王海洋) \orcidlink{0000-0001-7167-6110}}
\caltech
\burke

\title{Recoil geometry determines electromagnetic counterparts from \\
     supermassive black hole merger remnants}

\begin{abstract}
    Merging binary black holes embedded in gaseous environments, such as
    supermassive black hole binaries following gas-rich galaxy mergers, are
    promising sources of multi-messenger transients in the upcoming age of
    space-based gravitational wave detections. In case a gravitational radiation
    recoil is imparted to the merger remnant, subsequent interactions between
    the recoiled black hole and its circumbinary disk may lead to unique
    post-merger electromagnetic counterparts. We present the first general
    relativistic magnetohydrodynamic simulations of a recoiling black hole
    interacting with a magnetically arrested circumbinary disk the evolution of
    which has been consistently tracked through the inspiral phase. We show that
    the post-merger accretion dynamics, depending on the recoil geometry,
    exhibits qualitatively disparate jet and disk behavior. For recoils
    perpendicular to the disk, the inner disk remains gravitationally bound and
    sustains relativistic jets, while in-plane recoils lead to copious shock
    heating and potential jet quenching for black holes directly colliding with
    the disk. Oblique recoils, on the other hand, produce intermittent outbursts
    from jet-disk interactions owing to the tilt introduced in the accretion
    disk.
    Multi-wavelength monitoring of these electromagnetic counterparts, in
    conjunction with the coincident gravitational wave detection, will be able
    to aid in characterizing the physical conditions of the merger environment.
\end{abstract}

\maketitle
\end{CJK*}

\section*{Introduction}

Black holes (BHs) accreting and merging in gaseous environments can source
gravitational waves and coincident electromagnetic (EM) counterparts. Binaries
of supermassive black holes (SMBHs) \cite{Begelman:1980vb,Milosavljevic:2001vi}
are a primary target for low-frequency future gravitational wave detectors such
as Laser Interferometer Space Antenna \cite{LISA:2022yao} as well as a promising
source of EM counterparts. The nanohertz gravitational wave background recently
detected by pulsar timing arrays \cite{NANOGrav:2023hfp} also provides strong
evidence for a cosmological population of inspiraling SMBH binaries in a more
massive mass range. Several SMBH binary candidates have been identified to date
via EM observations [e.g.
\citealp{Sillanpaa:1988,Valtonen:2008nat,Graham:2015nat,Rodriguez:2006,Komossa:2003ngc,ONeill:2021swa,Kiehlmann:2024fji,Readhead:2025ihk,Molina:2025yut,Tremblay:2025fgk},
\citealp[see also][]{DOrazio:2023rvl}], where the precise physical conditions at
merger remain an open question \cite{Tiede:2025llq}. SMBH binaries formed in
galaxy mergers, particularly gas-rich ones, are believed to be surrounded by a
circumbinary disk (CBD) susceptible to dynamical perturbations from the evolving
binary \cite{Lai:2022ylu}. As a natural result of such astrophysically `wet'
environment, diverse mechanisms of EM counterparts to merging SMBH binaries
across a wide range of spatial and temporal scales have been suggested \cite[see
e.g.][for comprehensive discussions]{Bogdanovic:2021aav,DOrazio:2023rvl}.

Post-merger EM counterparts (afterglows) of a SMBH merger, in particular, not
only encodes the information on the remnant SMBH and its surroundings, but are
critical for source localizations and follow-up EM observations. Proposed
afterglow models include viscous refilling of the circumbinary disk
\cite{Milosavljevic:2004cg,Tanaka:2009iy,Shapiro:2009uy}, shock dissipation
within the disk triggered by gravitational mass loss
\cite{Milosavljevic:2004cg,ONeill:2008sat,Megevand:2009yx,Corrales:2009nv,Bode:2011pqa,Rosotti:2012yz},
or relativistic jets launched from the spinning remnant BH
\cite{Yuan:2021jjt,2018ASPC..517..781R}.

When a BH binary is asymmetric or individual components are spinning,
anisotropic gravitational wave emissions during the merger can impart a recoil
(`kick') to the remnant
\cite{Peres:1962zz,Bekenstein:1973zz,Fitchett:1983qzq,Baker:2006vn,Campanelli:2007cga,Gonzalez:2006md}.
In a merger of SMBHs, such a recoil can lead to a range of astrophysical
outcomes including spatially or kinematically displacing the SMBH from the
galaxy
\cite{Merritt:2004xa,Madau:2004st,Loeb:2007wz,Bonning:2007vt,Volonteri:2008gj,Komossa:2008as,Guedes:2010tk,Blecha:2015baa,2018ASPC..517..763B}
(see e.g.,
\cite{Shen:2019ssp,Ward:2020xji,Uppal:2024qxj,Barrows:2025,vanDokkum:2025bah,Islam:2026sjl}
for recent search campaigns),
tidal disruption events
\cite{Komossa:2008ye,Stone:2010sr,Stone:2011qa,Li:2012gn}, dynamical imprints on
stellar clusters
\cite{Merritt:2008kg,OLeary:2008tpn,OLeary:2011ykg,Akiba:2021,Akiba:2023ami,Bright:2024soo,Akiba:2025},
or feedback to the host galaxy
\cite{Devecchi:2009,Sijacki:2010tk,Boylan-Kolchin:2004fnd,Gualandris:2007nm,Blecha:2010dq}.
The instantaneous injection of a massive kinetic energy onto the remnant SMBH
and a consequent response of the CBD can power luminous EM afterglows
\cite{Milosavljevic:2004cg,Bode:2011pqa}. Numerical simulations show
{spiral-shaped} density enhancements and shocks develop in the perturbed CBD
\cite{Lippai:2008fx,Megevand:2009yx,Corrales:2009nv,Zanotti:2010xs,Zanotti:2011xq,Meliani:2016rwn},
dissipative luminosity of which can reach $\approx 10^{43}\,{\rm erg\,s^{-1}}$
rising over months to years for a $10^6 M_\odot$ SMBH
\cite{Rossi:2009nk,Corrales:2009nv,Zanotti:2010xs,Ponce:2011kv}. Such EM flares
are thought to peak at UV or soft X-ray energy band, while surrounding gas and
dust may reprocess them into infrared emissions
\cite{Schnittman:2008ez,Anderson:2009fa}.

A consistent modeling of the gaseous SMBH merger, due to its highly nonlinear
nature, requires numerical calculations in full general relativity. Along with
recent advances in computational techniques, the volume of the numerical studies
on accreting SMBH binaries has been rapidly growing in literature
\cite{Bode:2009mt,Farris:2009mt,Bode:2012,Farris:2011vx,Giacomazzo:2012iv,Farris:2012ux,Gold:2013zma,Gold:2014dta,Bowen:2016mci,Kelly:2017xck,Bowen:2017oot,Khan:2018ejm,Bowen:2019ddu,Armengol:2021shd,Cattorini:2021elw,Combi:2021dks,Cattorini:2022tvx,Paschalidis:2021ntt,Gutierrez:2021png,Bright:2022hnl,Ruiz:2023hit,Avara:2023ztw,Fedrigo:2023rvd,Cattorini:2023akr,Ressler:2024tan,Manikantan:2024giq,Ennoggi:2025nht,Tiwari:2025imm,Manikantan:2025afy,Chan:2025ers,Ennoggi:2025ijc,Manikantan:2025eqa,deSimone:2025cqg,Ressler:2025fti,Jiang:2026lop}.
Meanwhile, a detailed investigation on the recoiled SMBH in the post-merger
phase has remained relatively elusive. As demonstrated in early studies
\cite{Lippai:2008fx,Megevand:2009yx,Rossi:2009nk,Ponce:2011kv}, recoil
kinematics of the remnant BH relative to the disk can have profound impacts on
post-merger EM counterparts.

However, all prior studies dedicated to a recoiled SMBH interacting with its CBD
have employed collisionless or purely hydrodynamic approaches and did not
account for magnetic fields in the disk, crucially lacking the ability to
capture relativistic outflows from the BH or its interaction with magnetized
environments. Cosmological magnetohydrodynamics (MHD) simulations
\cite{Hopkins:2024forg1,Hopkins:2024forg2,Guo:2024mag} suggest that accretion
disks around individual SMBHs prior to merger could be generically strongly
magnetized, and recent galactic-scale simulations showed the formation of a
strongly magnetized CBD around SMBH binaries \cite{Wang:2025mit}. Furthermore,
recent long-term MHD simulations over hundreds of orbits
\cite{Most:2024qus,Most:2024onq,Wang:2025yco} found the SMBH binary, as it
approaches merger, can be immersed in a common magnetized funnel surrounded by a
strongly magnetized CBD resembling the magnetically arrested disk (MAD) state
\cite{Narayan:2003by,Igumenshchev:2007bh} of an isolated, accreting BH. The
presence of these strong magnetic fields enables novel nonthermal EM signatures
which have been appreciated only recently \cite{Most:2024onq,Ressler:2024tan}
and remained fully unexplored in terms of recoil-driven transients.

In this paper, we present general relativistic MHD (GRMHD) simulations of a
recoiled, spinning BH accreting from and disrupting a magnetically arrested CBD.
Building upon a realistic {CBD} configuration evolved over hundreds of orbits in
the pre-merger phase \cite{Most:2024qus,Most:2024onq}, our study marks the first
systematic investigation in relativistic MHD framework on how the kinematics of
the recoiled SMBH affects prompt nonthermal and thermal EM counterparts.

\section*{Methods}

We perform numerical ideal GRMHD simulations using the \athenak~code
\cite{Stone:2026}. A snapshot from a Newtonian MHD simulation of an accreting BH
binary \cite{Most:2024onq}, which has tracked the evolution of a magnetically
arrested CBD up to the late-inspiral decoupling phase
\cite{Armitage:2002uu,Dittmann:2023dss}, is imported as initial data. Our
simulations are performed in the rest frame of the remnant BH using a fixed Kerr
background spacetime, and the recoil of the BH is included by adding an equal
and opposite velocity offset to the imported MHD matter profile. Gravitational
mass loss in the merger is ignored and the BH mass is set to be equal to the
mass of the original binary. See the Supplemental Material for the full
procedure of preparing initial data. Evolution of the gas is modelled with an
ideal gas equation of state. The spin of the remnant BH is assumed to be aligned
with the disk midplane, and we adopt the dimensionless spin magnitude $\chi =
0.9$.

The gravitational cross section of the recoiling BH is given by the
Bondi-Hoyle-Lyttleton accretion radius \cite{Hoyle1939,Edgar:2004mk} $R_a =
2GM_{\rm BH}/v_r^2$, which we aim to resolve in this work. A realistic recoil
speed\footnote{Non-spinning BH binaries can have a recoil up to $\lesssim 200 \,
{\rm km \, s^{-1}}$ \cite{Gonzalez:2006md}. Mergers of spinning binaries
\cite{Campanelli:2007ew,Koppitz:2007ev,Herrmann:2007ac,Baker:2008md,Lousto:2008dn},
especially if the individual spins are arranged into a specific configuration
(`superkick' \cite{Campanelli:2007cga,Gonzalez:2007hi} or `hang-up'
\cite{Lousto:2011kp,Lousto:2012su}), can lead to a higher recoil speed $\lesssim
5000 {\rm \,km\,s^{-1}}$. See, however,
\cite{Bogdanovic:2007hp,ColemanMiller:2013jrk} on the potential spin alignment
of the SMBH binary during galactic mergers.} of a merger remnant, however,
yields $R_a/r_g = (v_r/c)^{-2} \gtrsim 10^4$ where $r_g \equiv GM_{\rm BH}/c^2$
is the gravitational radius of the BH, demanding an extreme computational cost
to resolve both length scales in a single computational domain. In this regard,
we adopt the recoil speeds computationally affordable yet maintaining a fair
scale separation: $10^2 \lesssim R_a/r_g \lesssim 10^3$. Specifically, we
consider the BH recoil speed $v_r = 0.10c$ and $v_r = 0.05c$, but put more
emphasis on the latter in discussing the results. To investigate the effect of
recoil geometry, we consider three different recoil angles: vertical
(perpendicular to the disk midplane), oblique (45 degrees), and horizontal
(in-plane). For horizontal recoils, we additionally investigate slower recoil
speeds $v_r=0.03c$ and $v_r=0.01c$. Simulation parameters are summarized in
Table~\ref{tab:parameters}.

\begin{table}
\centering
\caption{List of the simulation parameters.}
\label{tab:parameters}
\setlength{\tabcolsep}{5pt}
\begin{tabular}{c|c|l}
    \hline
    Label & Recoil velocity ($c$) & Description \\
    \hline\hline
    \texttt{v010$\uparrow$} & (0, 0, 0.10) & \multirow{2}{*}{vertical (out-of-plane) recoil} \\
    \texttt{v005$\uparrow$} & (0, 0, 0.05) & \\
    \hline
    \texttt{v010$\nearrow$} & (0, 0.071, 0.071) & \multirow{2}{*}{oblique recoil} \\
    \texttt{v005$\nearrow$} & (0, 0.035, 0.035) & \\
    \hline
    \texttt{v010$\rightarrow$} & (0.10, 0, 0) & \multirow{4}{*}{horizontal (in-plane) recoil} \\
    \texttt{v005$\rightarrow$} & (0.05, 0, 0) & \\
    \texttt{v003$\rightarrow$} & (0.03, 0, 0) & \\
    \texttt{v001$\rightarrow$} & (0.01, 0, 0) & \\
    \hline
\end{tabular}
\end{table}

\begin{figure*}
\center
\includegraphics[width=\linewidth, page=1]{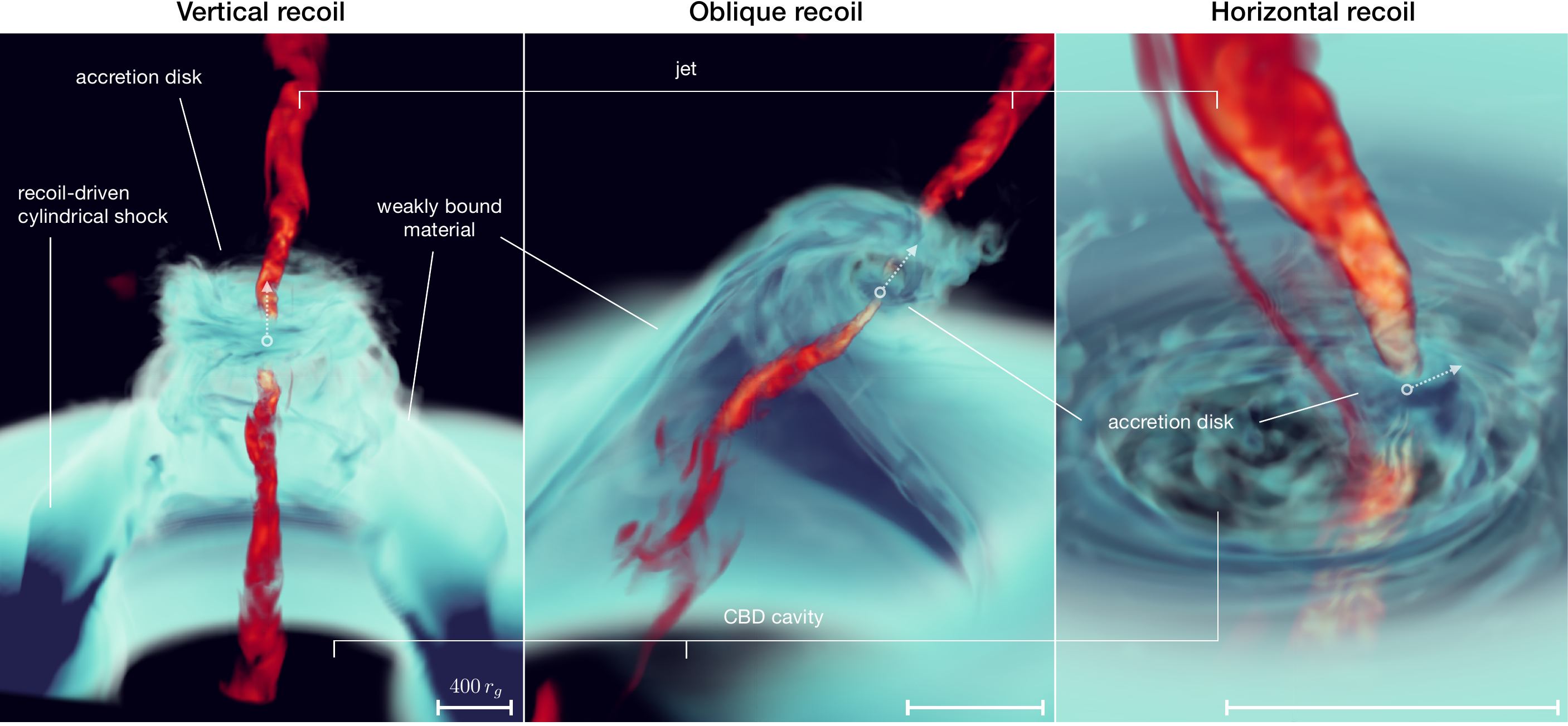}
\caption{Three-dimensional view of the rest-mass density (white-blue) and the
    relativistic magnetization parameter $\sigma$ (red-yellow) from the
    simulations with a vertical recoil at $v_rt = 1934 r_g$ (left), an oblique
    recoil at $v_r t = 1354 r_g$ (center), and a horizontal recoil at $v_r t =
    166 r_g$ (right), all for the recoil speed $v_r = 0.05c$. Location and
    recoil direction of the BH are shown with a dotted arrow. Key features are
    annotated with plain texts.
    }
\label{fig:volume-rendering}
\end{figure*}

\begin{figure*}
    \center
    \includegraphics[width=\linewidth, page=1]{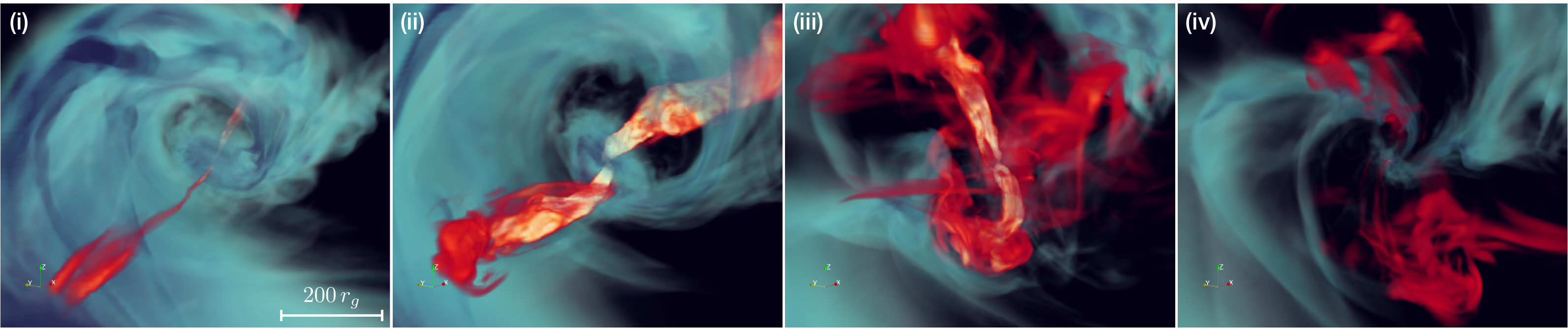}
    \caption{A zoom-in view of an jet outburst episode in the obliquely recoiled
        model (\texttt{v005$\nearrow$}). Each panels have the same viewing angle
        as in the center panel in Fig.~\ref{fig:volume-rendering}, and
        correspond to simulation times $t = $ (i) $2.8\e{4} r_g c^{-1}$, (ii)
        $3.9\e{4} r_g c^{-1}$, (iii) $4.6\e{4} r_g c^{-1}$, (iv) $5.0\e{4} r_g
        c^{-1}$.}
    \label{fig:oblique-jet-disk-interaction}
\end{figure*}

\begin{figure}
    \center
    \includegraphics[width=\linewidth]{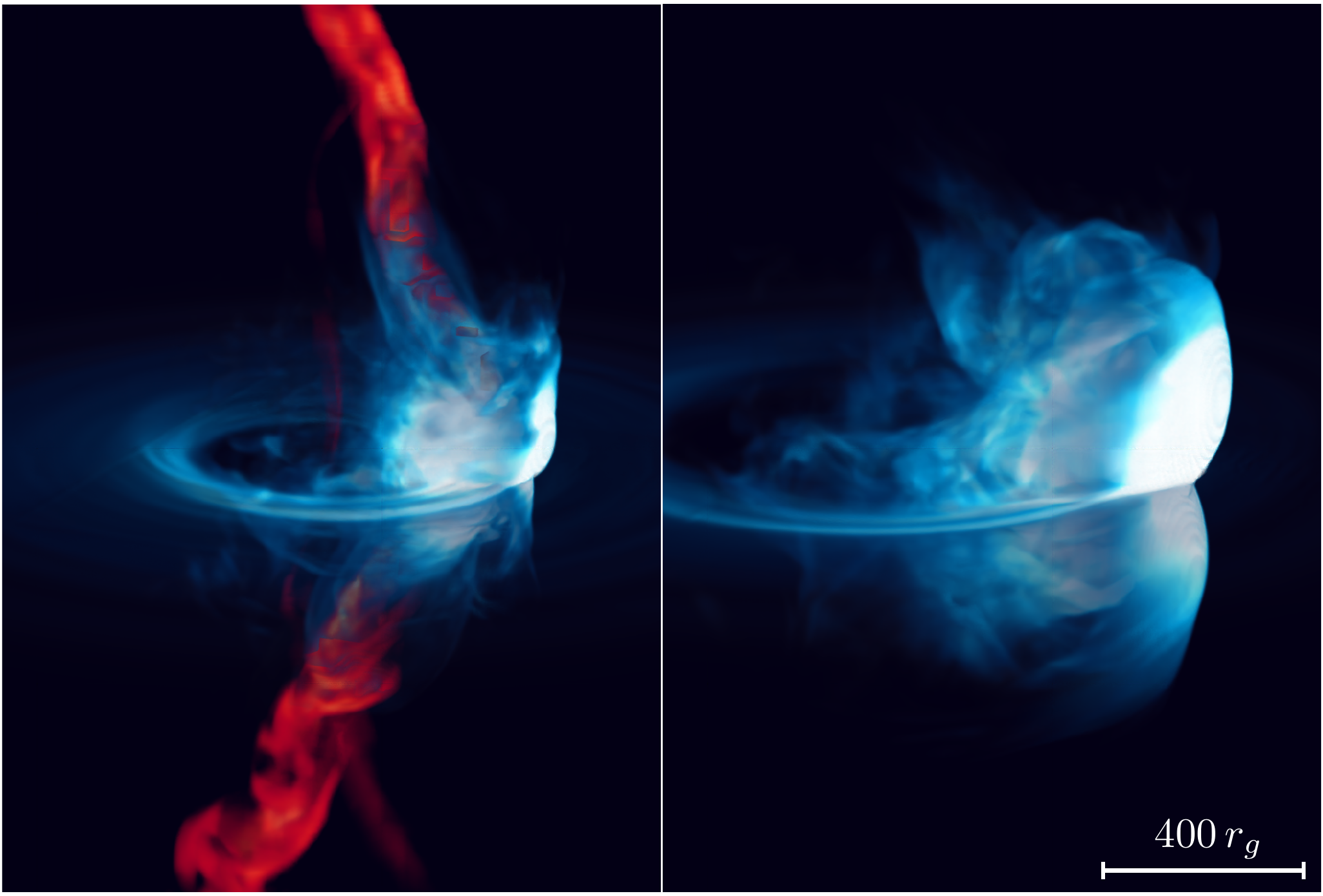}
    \caption{Magnetization parameter $\sigma$ (red) and the internal energy
        density (bright blue) in the \texttt{v005$\rightarrow$} model at $v_rt =
        332 r_g$ (left) and $v_rt = 663 r_g$ (right), showing the jet and the
        distribution of the shock-heated gas. The recoiled BH (not visible here)
        is moving to the right side of the figure.
        }
    \label{fig:heating}
\end{figure}
    
\section*{accretion flow and jet dynamics}

Owing to large supply of vertical magnetic fluxes in the magnetically arrested
CBD cavity, the remnant BH quickly saturates to the MAD state and launches
relativistic jets along its spin \cite{Blandford:1977ds,Tchekhovskoy:2011zx}.
The subsequent evolution of the accretion flow and its interaction with the jet
exhibit a strong dependency on the recoil angle, as we describe in detail below.
Fig.~\ref{fig:volume-rendering} shows the global morphology of the accretion
flow and jets for the models with the recoil speed $v_r=0.05c$ and different
angles (\texttt{v005$\uparrow$}, \texttt{v005$\nearrow$},
\texttt{v005$\rightarrow$}).

A vertically recoiled BH is ejected from the disk midplane along with the inner
part of the CBD that remains gravitationally bound, the radial extent of which
roughly agrees with the accretion radius $R_a = 2GM_{\rm BH}/v_r^2$. The jet and
the bound accretion disk (i.e. $r\lesssim R_a$, hereafter `accretion disk' to
disambiguate from the CBD) remains fairly intact since the recoil {is purely
out-of-plane}. A diffuse ejecta of weakly bound material expanding near the
radius $r\gtrsim R_a$, demonstrated in previous numerical studies
\cite{Lippai:2008fx,Shields:2008va,Rossi:2009nk}, exhibits more turbulent
substructures in our simulations due to MHD effects. A cylindrical shock
dividing the diffuse ejecta and unperturbed CBD expands outward, plowing through
the CBD and depositing a fraction of the shocked gas into the ejecta. While the
recoiled BH maintains a steady jet launching, we observe a gradual decrease in
the mass accretion rate over time as the gas in the accretion disk is depleted.

An oblique recoil, instead, instantaneously excites the misaligned component in
the angular momentum of the accretion disk {in the BH frame}. The accretion
midplane of the incoming gas flow, channeled through a streamer connected to the
outer CBD, is highly warped and tilted (see the center panel of
Fig.~\ref{fig:volume-rendering}). A recoil-driven cylindrical shock, similar to
the one in the vertical recoil models, is launched with its strength enhanced
toward the projected direction of the recoil. In
Fig.~\ref{fig:oblique-jet-disk-interaction}, we illustrate the interplay between
the jet and the tilted accretion flow. The jet is originally aligned with the BH
spin, but its propagation direction is progressively tilted and steered by the
incoming accretion flow \cite{Liska:2017alm}. Intermittently when the gas influx
is low, the jet overwhelms the gas pressure and forcefully aligns the inner
accretion disk with the BH spin \cite{McKinney:2012wd}. This process severely
disrupts the accretion disk, effectively self-destructing its own gas supply,
and in turn weakens the jet. Then, the next episode of this accretion-disruption
cycle continues. A very similar behavior has also been reported in a recent
study on the accretion of highly misaligned disks \cite{Chatterjee:2023ber}. As
a result, oblique recoil models show the highest degree of complexity and
variability in the region $r\lesssim R_a$, and comparisons between
\texttt{v010$\nearrow$} and \texttt{v005$\nearrow$} do not exhibit a clear
trend. Violent jet-disk interactions described above can rapidly dissipate the
jet Poynting flux and trigger EM transients. These episodic outbursts can also
leave highly inhomogeneous traces of BH feedback, characterized by disconnected
segments of magnetized outflows stemming from a common linear recoil path.

In the horizontally (in-plane) recoiled models, the remnant BH directly collides
with the CBD and is subject to a parallel inflow of abundant gas. A strong bow
shock forms ahead of the BH and expands over time, resembling the early phase of
the Bondi-Hoyle-Lyttleton accretion
\cite[e.g.][]{Shima:1985,Edgar:2004mk,Ruffert:1994} (see also
Ref.~\cite{Kaaz:2022dsa,Gracia-Linares:2023yxw,Kim:2024zjb} for recent MHD
simulations). In Fig.~\ref{fig:heating}, we render the gas internal energy
density in the \texttt{v005$\rightarrow$} model, revealing the distribution of
shock-heated material. A higher gas mass density along the equatorial plane
gives rise to a concave kink at the middle of the bow shock. The jet from the BH
is bent to the opposite direction of the recoil by the ram pressure of the
incoming gas \cite{Kaaz:2022dsa}. We find the jet launching is quenched after $t
\sim 10^4 r_g c^{-1}$ in the \texttt{v005$\rightarrow$} model, where
\texttt{v010$\rightarrow$} and \texttt{v003$\rightarrow$} models show early and
later jet quenching time, respectively. Once the jet is switched off, mass
accretion rate rapidly increases and the subsequent accretion proceeds more
closely in the matter pressure-dominated (hydrodynamic) regime. An undulating
wake pattern is formed along the trailing path behind the BH (see
Fig.~\ref{fig:cavity-reconnection}), similar to the one previously seen in a 2D
hydrodynamic simulation \cite{Rossi:2009nk}.

\section*{Electromagnetic signatures}

The gravitationally bound material comoving with the BH continues to fuel
accretion and can source prolonged activity as an off-center active galactic
nuclei (AGN) \cite{Loeb:2007wz}. The relativistic jet and the hot accretion disk
can continue to source nonthermal radio/X-ray and thermal UV/optical emissions,
respectively. Unless the recoil direction is closely aligned to the CBD
midplane---for which a substantial amount of gas is still available along the
recoiling trajectory---jet launching and accretion will cease once the finite
amount of gas within the BHs sphere of influence is depleted. The lifetime of
this `mini-AGN' ejected off from the CBD can be estimated as $\tau_{\rm acc}
\equiv M_b/\dot{M}$,\footnote{Viscous spreading of the disk can extend its
lifetime by a factor of a few or more. A self-similar solution of an isolated
$\alpha$-disk predicts $\dot{M}(t) \propto t^{-19/16}$
\cite{Pringle1974,Cannizzo:1990}} where $M_b$ is the total mass of the
gravitationally bound material and $\dot{M}$ is the mass accretion rate.
In the \texttt{v010$\uparrow$} model, we see that $\tau_{\rm acc}$ relaxes to
$\approx 10^6 r_g c^{-1}$ by the end of the simulation (see Supplemental
Material). Taking this as a fiducial data point, we postulate the following
scaling estimate
\begin{equation} \label{eq:accretion lifetime scaling}
    \tau_{\rm acc} \sim 16
        \left(\frac{M_{\rm BH}}{10^8  M_\odot}\right)
        \left(\frac{v_r}{0.1c}\right)^{-2p}
        \, {\rm yrs}  \, .
\end{equation}
with an assumption that $\tau_{\rm acc} \propto M_b \approx M_{\rm disk}(r<R_a)$
and a power-law scaling $M_{\rm disk} (r < R_a) \propto (R_a)^p$. The exponent
$p$ depends on the detail of the SMBH disk model; for instance, the standard
$\alpha$-disk model \cite{Shakura:1972te} in a steady state gives $p=7/5$
\cite{Goodman:2003sf,Loeb:2007wz}. Assuming a recoil velocity $v_r = 1000\, {\rm
km\,s^{-1}}$ and $p=7/5$, we get $\tau_{\rm acc} \sim 10^8$ years for a $10^8
M_\odot$ SMBH. We remark that this extrapolation breaks down at small recoil
speeds at which (i) $M_{\rm disk}(r<R_a)$ becomes comparable to $M_{\rm BH}$,
(ii) stellar velocity dispersion in the galactic core is comparable to $v_r$, or
(iii) $R_a$ far exceeds the size of the CBD \cite{Madau:2004st,Loeb:2007wz}. The
conditions outlined above generally require a fast recoil $\gtrsim 1000\, {\rm
km\,s^{-1}}$. A more realistic estimate on the accretion lifetime $\tau_{\rm
acc}$ should account for galactic environment at $r\gtrsim 10^4 r_g$, which
should be addressed in future studies.

\begin{figure}
    \center
    \includegraphics[width=\linewidth]{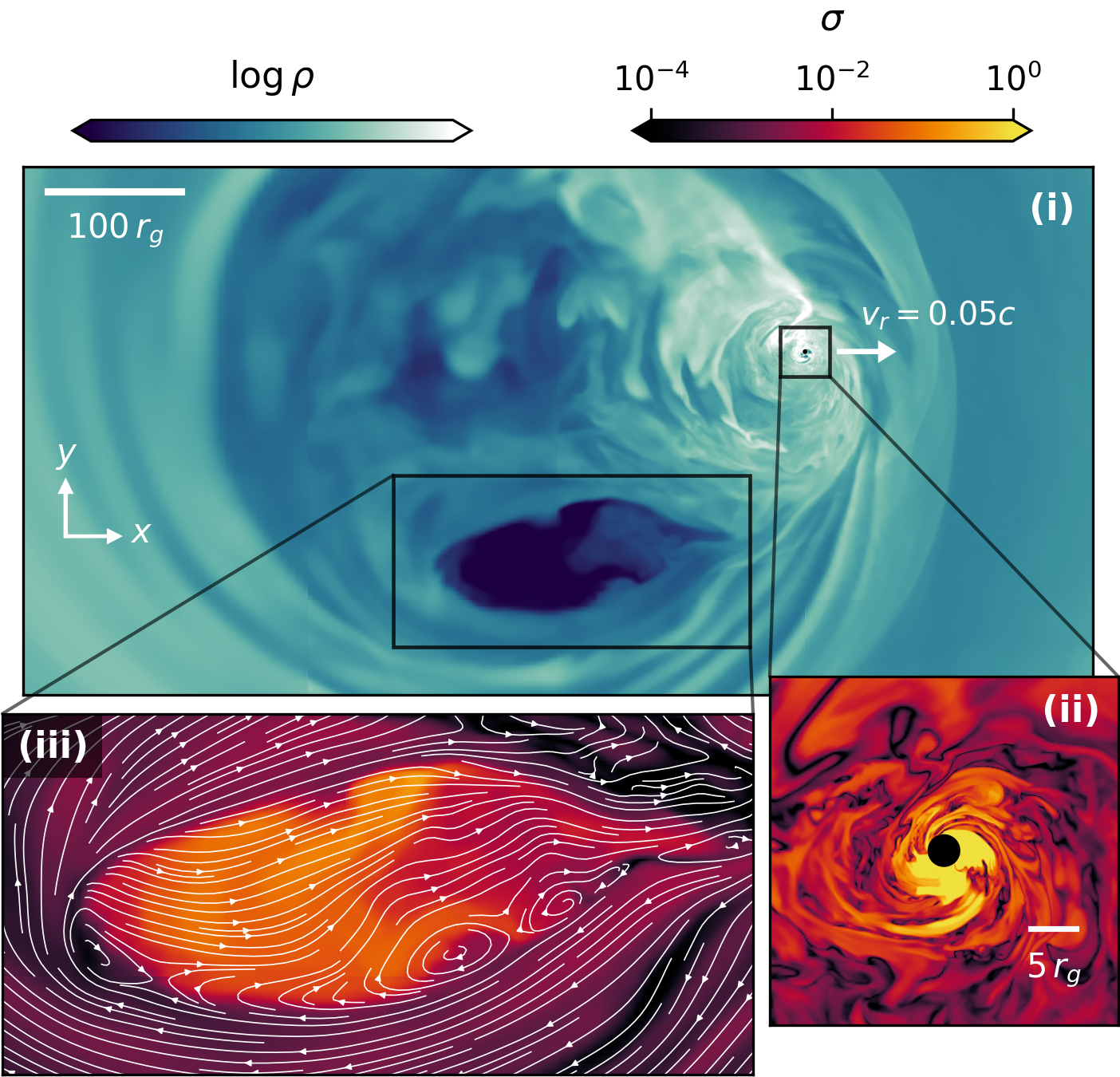}
    \caption{(i) Rest-mass density distribution on the equatorial plane in the
        \texttt{v005$\rightarrow$} model at $v_rt=207r_g$. (ii) The recoiling BH
        is in the MAD state and surrounded by a strongly magnetized accretion
        flow ($\sigma \gtrsim 1$). (iii) A fraction of the CBD cavity survives
        as a magnetized low-density blob, then being advected and sheared by
        ambient gas flow. The in-plane magnetic field lines (white solid curves)
        reveal multiple X-points, the site of active magnetic reconnection.}
    \label{fig:cavity-reconnection}
\end{figure}

An additional source of nonthermal flares, particularly in the case of
horizontal recoil, is the magnetic flux tube that breaks off from the CBD
cavity---a feature unique to the MAD regime. We show an example in
Fig.~\ref{fig:cavity-reconnection}. This region is likely to provide for
nonthermal particle acceleration owing to (1) high initial magnetization
reaching $\sigma = b^2/\rho c^2 \sim 1$, where $b^2$ is the magnetic energy
density and $\rho$ is the rest-mass density, and (2) magnetic field topology
conducive to efficient energization, including a somewhat turbulent interior
\cite[e.g.][]{Vega:2024pkx} and relativistic asymmetric reconnection at its
boundary \cite{Mbarek:2021vja}. In addition, the interface between the flux tube
and the surrounding gas flow is subject to the magnetic Rayleigh-Taylor
instability \cite{Zhdankin:2023wch}, which can further enhance dissipation and
nonthermal particle {acceleration}. Near the BH, where the flux tube remains
strongly magnetized ($\sigma \gtrsim 1$), these processes may generate TeV and
X-ray flares \cite{Porth:2020txf,Ripperda:2021zpn,Hakobyan:2022alv}, whereas as
the tube drifts away from the BH and its magnetization decreases ($\sigma
\lesssim 10^{-2}$), the emission may shift toward lower energies such as
infrared flares \cite{Dexter:2020cuv,Most:2024onq}. For horizontal recoils,
where the BH can continue to accrete a substantial supply of magnetized gas,
quasi-periodic magnetic flux eruptions \cite{Chatterjee:2022mxg} may repeatedly
release similar highly magnetized flux tubes downstream \cite{Kim:2024zjb},
sustaining intermittent nonthermal flare activities. Although we postpone a
detailed analysis of whether synchrotron or inverse Compton emission dominates
within these structures, their high magnetization, strong dissipation, and
access to UV seed photons from the accretion disk suggest that they could
temporarily behave as transient BH X-ray coronae, upscattering the UV photons
into hard X-rays.

\begin{figure}
    \center
    \includegraphics[width=\linewidth]{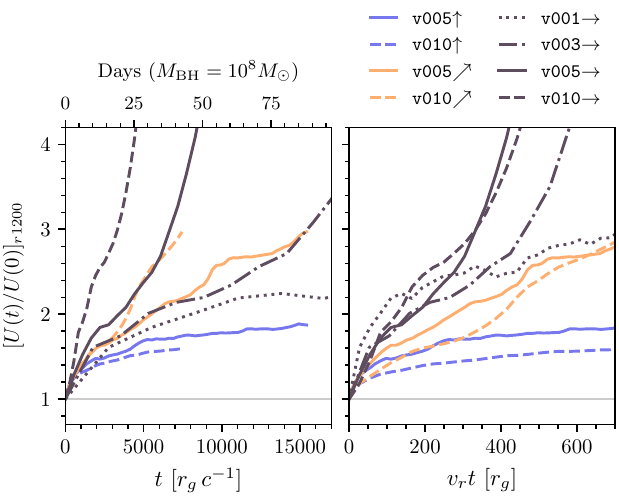}
    \caption{Total gas internal energy $U(t)$ contained within a spherical
        volume $r < 1200 r_g$ centered to the original circumbinary disk,
        normalized with its initial value. Horizontal scales are time (left) and
        the recoil distance (right).}
    \label{fig:energetics}
\end{figure}

In order to probe the energy dissipation in the perturbed disk, we compute the
total gas internal energy $U = \int e u^0 \sqrt{-g} \, d^3x$ within a spherical
volume $r < 1200r_g$ comoving with the original center of the CBD. As a simple
diagnostic, total internal energy provides an estimate for the energy budget
associated with thermal photon emission from the disk. Fig.~\ref{fig:energetics}
shows, for all models, the time evolution of $U(t)$ normalized with its initial
value. The left panel is plotted in simulation time, while the right panel shows
the same data with the time axis normalized to the coordinate recoil distance.

For horizontal and oblique recoil angles, data from different recoil speeds
roughly overlap with each other in the right panel of Fig.~\ref{fig:energetics},
indicating that the collision of the BH (plus the bullet of gas tightly bound to
it) with the disk is a primary heating mechanism. Fig.~\ref{fig:heating}
illustrates this recoil-driven shock heating in the CBD. Compared to horizontal
recoils, oblique recoils show a lower overall heating rate due to their
collision path misaligned with the disk midplane. A notable exception from this
trend is the \texttt{v001$\rightarrow$} model, and this anomaly comes from the
recoil being initially subsonic. The sound speed of the CBD in our setup drops
below $0.01c$ only at the radius $r\gtrsim 200r_g$, and the bow shock is
observed to develop only after $v_r t \gtrsim 300 r_g$. As the BH continues to
fly toward the cold outer part of the CBD, however, the recoil Mach number could
increase and the shock heating can be enhanced to give a secondary rise in the
thermal luminosity. The case of the \texttt{v001$\rightarrow$} model suggests
that early thermal afterglows from the disk can be kinematically ambiguous, as a
slow horizontal recoil may produce an EM signature similar to a faster, oblique
recoil. However, as discussed earlier, the recoil angle greatly alters the jet
intermittency and variability of the thermal emission from the innermost part of
the accretion disk, which can assist break the degeneracy.

Vertically recoiled cases exhibit a brief surge in $U(t)$ followed by a gradual
rise as the cylindrical shock is launched outward (see the first panel of
Fig.~\ref{fig:volume-rendering}). Heating is much lower than other models
because of a relatively weak density enhancement in the disk
\cite{Lippai:2008fx,Megevand:2009yx}. Meanwhile, it has been suggested that a
fallback of the weakly bound ejecta after about a few $GM_{\rm BH}/v_r^3$ and
its collision with the accretion disk can rapidly release a large amount of
energy $\approx M_{\rm ej} v_r^2/2$ in shocks that can power bright and long
soft-X ray emissions \cite{Shields:2008va}. Our simulations could only capture
an initial expansion of the ejecta.

If the mass ratio and orbital inclination of a black hole binary could be
determined from a concurrent gravitational-wave detection during the inspiral
phase, the recoil speed and angle can be predicted using analytic fits or
surrogate recoil models
\cite{Lousto:2009mf,Healy:2016lce,Gerosa:2018qay,Varma:2018aht,Islam:2025drw}
calibrated to numerical relativity simulations. When combined with the
post-merger EM signature, this multi-messenger observation would be able to
constrain the {physical conditions} of the merger environment---such as the
local recoil Mach number or the orbital misalignment between the binary and the
circumbinary disk---that might remain unresolved or obscured {in direct
observations.}

Future work incorporating cooling effects
\cite[e.g.][]{Wang:2025yco,Ennoggi:2025nht} and radiative transport
\cite[e.g.][]{Tiwari:2025imm,Chan:2025ers,Zhang:2025uug}, crucial to regulating
mass accretion onto the BH and thermal evolution of the perturbed disk, would be
able to provide a more realistic diagnostics of prompt electromagnetic emissions
from gaseous SMBH mergers.

\vspace{1em}

We are grateful to Manuela Campanelli, Alexander Dittmann, Jeremy Goodman,
Matthew Graham, Aretaios Lalakos, Yuri Levin, Rostom Mbarek, Eliot Quataert,
Daniel Stern, and Hengrui Zhu for insightful comments and discussions. Y.K.
acknowledges support by the Sherman Fairchild Foundation, by the National
Science Foundation through the grants No. PHY-2309211, No. PHY-2309231, and No.
OAC-2209656 at Caltech, and by a joint fellowship at the Princeton Center for
Theoretical Science and the Princeton Gravity Initiative. ERM and HYW
acknowledge support by the U.S. National Science Foundation under grant
NSF-AST2508940.
Simulations were performed on NERSC Perlmutter cluster through allocation m4575
and m5081. Additional simulations were carried out on AMDs AI \& HPC cluster.
This research used resources of the National Energy Research Scientific
Computing Center, which is supported by the Office of Science of the U.S.
Department of Energy under Contract No. DE-AC02-05CH11231.
Figures in this article were produced using \texttt{matplotlib}
\cite{matplotlib}, \texttt{numpy} \cite{numpy}, \texttt{scipy} \cite{scipy},
\texttt{Paraview} \cite{paraview:2005,paraview:2015}, and \texttt{yt}
\cite{2011ApJS..192....9T}.

\onecolumngrid

\bibliography{references}

\newpage
\section*{Appendix}

\section{preparation of numerical initial data}
\label{sec:initial data}

Here we describe the conversion procedure of a simulation snapshot from
\citet[][hereafter \citetalias{Most:2024onq}]{Most:2024onq} into the numerical
GRMHD initial data used in this paper. The Newtonian MHD simulation of an
accreting BH binary in \citetalias{Most:2024onq} was performed with the
\athenak~code, and we use a \athenak~checkpoint output near the moment of the
merger of the binary, specifically at $t = 1225 (a_0^3/GM)^{1/2}$ in their
setup.

The Newtonian simulation in \citetalias{Most:2024onq} was performed in the units
$a_0=GM=1$, where $a_0$ is the initial orbital separation and $M$ is the total
mass of the binary. The CBD was assumed to be non self-gravitating, and the unit
of mass is scalable by picking an arbitrary mass density $\rho_0$. Conversion
between physical values and code values are given as follows:
\begin{subequations} \label{eq:unit-conversion-newtonian}
\begin{align}
  x^i  &= a_0  \bar{x}^i\,, \\[1ex]
  v^i  &= \sqrt{\frac{GM}{a_0}} \,\bar{v}^i \,,\\[1ex]
  \rho &= \rho_0\, \bar{\rho}\,, \\[1ex]
  e    &= \rho_0 \left(\frac{GM}{a_0}\right) \, \bar{e} \,,\\[1ex]
  B^i  &= \left(\frac{GM \rho_0}{a_0}\right)^{1/2} \bar{B}^i\,,
\end{align}
\end{subequations}
where $x^i$ is spatial coordinate, $v^i$ is fluid velocity, $\rho$ is mass
density, $e$ is internal energy density, and $B^i$ is the magnetic field. Code
values are denoted with barred variables on the right hand sides of
Eqs.~\eqref{eq:unit-conversion-newtonian}.

The GRMHD accretion simulation performed in this paper adopts the unit system $c
= GM_{\rm BH} = 1$, also with an arbitrary mass density scale $\varrho_0$. The
conversion rules between physical and code values are
\begin{subequations} \label{eq:unit-conversion-grmhd}
\begin{align}
  x^i  &= \frac{GM_{\rm BH}}{c^2} \, \hat{x}^i \,,\\[1ex]
  v^i  &= c \, \hat{v}^i\,, \label{eq:unit-conversion-grmhd-velocity} \\[1ex]
  \rho &= \varrho_0 \, \hat{\rho}\,, \\[1ex]
  e    &= (\varrho_0 c^2) \, \hat{e}\,, \\[1ex]
  B^i  &= (\varrho_0^{1/2} c) \, \hat{B}^i\,,
\end{align}
\end{subequations}
where the code values are denoted with hatted variables on the right hand sides.
Note that $r_g \equiv GM_{\rm BH}/c^2$ is the gravitational radius $r_g$ defined
in the main text of the paper.

Combining \eqref{eq:unit-conversion-newtonian} and
\eqref{eq:unit-conversion-grmhd}, we get the conversion rules of physical
quantities between the two code units:
\begin{align}
  \label{eq:conversion-spatial-coordinates}
  \hat{x}^i  &= \left(\frac{GM}{a_0 c^2}\right)^{-1}
                \left(\frac{M_{\rm BH}}{M}\right)^{-1} \, {\bar{x}}^i \,,\\[1ex]
  \label{eq:conversion-velocity}
  \hat{v}^i  &= \left(\frac{GM}{a_0 c^2}\right)^{1/2} \, \bar{v}^i \,,\\[1ex]
  \label{eq:conversion-mass density}
  \hat{\rho} &= \left(\frac{\rho_0}{\varrho_0}\right) \, \bar{\rho} \,,\\[1ex]
  \label{eq:conversion-energy density}
  \hat{e}    &= \left(\frac{\rho_0}{\varrho_0}\right)
                \left(\frac{GM}{a_0 c^2}\right) \, \bar{e} \,,\\[1ex]
  \label{eq:conversion-magnetic field}
  \hat{B}^i  &= \left(\frac{\rho_0}{\varrho_0}\right)^{1/2}
                \left(\frac{GM}{a_0 c^2}\right)^{1/2} \, \bar{B}^i\,,
\end{align}

Brief comments on the three rescaling factors appearing in
Eq.~\eqref{eq:conversion-spatial-coordinates}--\eqref{eq:conversion-magnetic
field}:
\begin{itemize}[leftmargin=2em, labelsep=1em]
\item The factor $GM/a_0 c^2$ describes how relativistic the orbital motion of
    the binary is, and was taken to be $GM/a_0 c^2= (27.16)^{-1}$ in
    \citetalias{Most:2024onq}, which we use for the mapping procedure.
\item The factor $M_{\rm BH}/M$ in Eq.~\eqref{eq:conversion-spatial-coordinates}
    can be used to account for the gravitational mass loss of the merger remnant
    by taking a specific value $M_{\rm BH} / M < 1$. The mass loss has been
    neglected in this work and was set to $M_{\rm BH} / M = 1$.
\item The choice of the mass density scales $\rho_0$ and $\varrho_0$ are both
    arbitrary, and can be taken to be $\rho_0/\varrho_0 = 1$ for simplicity. For
    practical purposes, an appropriate choice of $\rho_0 / \varrho_0$ can help
    rescale the code values $(\hat{\rho}, \hat{e}, \hat{B}^i)$ to be numerically
    more tractable in a new code unit.
\end{itemize}

In summary, we adopt $GM/a_0c^2 = (27.16)^{-1}$, $M_{\rm BH} / M = 1$, and
$\rho_0/\varrho_0 = 1$ in the rescaling
\eqref{eq:conversion-spatial-coordinates}--\eqref{eq:conversion-magnetic field}.
Now we describe the initial data mapping process in order.

\subsection{Domain and grid structure} \label{sec:amr}

First, grid refinement levels in the \citetalias{Most:2024onq} simulation data
is manipulated using the mesh refinement functionality of \athenak~code
\cite[see][]{Stone:2020,Stone:2026}. Prolongation and restriction operations
that are built in the code framework respect global conservation of the evolved
variables and divergence-free condition of magnetic fields during this
procedure. In general, preserving both conditions in high accuracy could be
challenging if the matter profile is interpolated onto a new set of grid points.
We essentially bypass this difficulty by recycling the computational domain
aside from adjusting refinement levels. Additional grid refinement is applied
near the coordinate origin while the regions far from the origin are coarsened,
if possible, to reduce the computational cost. Spatial coordinates of each grid
points are uniformly rescaled following
Eq.~\eqref{eq:conversion-spatial-coordinates}, then the Kerr metric in the
Cartesian Kerr-Schild coordinates is set as the background spacetime. The final
configuration of the grid has the domain size $[-2173r_g, 2173r_g]^3$ and the
finest resolution of 23 grid points per $r_g$ around the BH located at the
origin.

\subsection{Rescaling and mapping primitive variables} \label{sec:mapping}

The primitive variables of the Newtonian MHD solver used in
\citetalias{Most:2024onq}, which are the variables dumped into the checkpoint
file, are
\begin{equation} \label{eq:mhd-prim-newtonian}
    (\rho, v^i, e, B^i) \, ,
\end{equation}
where $\rho$ is the mass density, $v^i$ is the fluid velocity, $e$ is the
internal energy density, and $B^i$ is the magnetic field. The primitive
variables used in the (non-dynamical spacetime) GRMHD solver of \athenak~are
\begin{equation}
    (\rho, u^{i'}, e, \mathcal{B}^i)
\end{equation}
where $\rho$ is the rest mass density,
\begin{equation} \label{eq:prim-vel}
    u^{i'} = u^i + \beta^i u^0
\end{equation}
is normal-frame spatial velocity, $e$ is the internal energy density, and
\begin{equation} \label{eq:prim-B}
    \mathcal{B}^i = \dual{F}^{i0}  %
\end{equation}
is the coordinate frame magnetic field. In \eqref{eq:prim-vel} and
\eqref{eq:prim-B}, $u^\mu = (u^0, u^i)$ is the fluid 4-velocity, $\beta^i =
-g^{0i}/g^{00}$ is the shift vector in the standard 3+1 decomposition of the
spacetime metric, and $\dual{F}^{\mu\nu}$ is the dual electromagnetic field
tensor.

Code values of the Newtonian MHD primitive variables
\eqref{eq:mhd-prim-newtonian} are plugged in as the barred variables on the
right hand side of
Eq.~\eqref{eq:conversion-velocity}--\eqref{eq:conversion-magnetic field}. Let us
denote the resulting values of the hatted variables as
\begin{equation}
    (\hat{\rho}, \hat{v}^i, \hat{e}, \hat{B}^i)_{\rm Newt} \, ,
\end{equation}
based on which we need to assign the in-code GRMHD primitive variables
\begin{equation}
    (\hat{\rho}, \hat{u}^{i'}, \hat{e}, \hat{\mathcal{B}}^i)_{\rm GR} \, .
\end{equation}

Identification of the Newtonian primitive variables to their general
relativistic counterparts is ambiguous and there may exist several possible
choices; for example, Newtonian 3-velocity $v^i$ can be mapped to $u^i$ or
$u^i/u^0$, and so on. While keeping this caveat in mind, we describe our
approach below.

\begin{itemize}[leftmargin=2em, labelsep=1em]
\item The rest mass density and the internal energy density are used without any
    modification i.e. $\hat{\rho}_{\rm GR} \leftarrow \hat{\rho}_{\rm Newt}$ and
    $\hat{e}_{\rm GR} \leftarrow \hat{e}_{\rm Newt}$.
\item We regard the rescaled Newtonian velocity $(\hat{v}^i)_{\rm Newt}$ as
    $(\hat{u}^i)_{\rm GR}$, spatial components of the 4-velocity. Then,
    \begin{enumerate}[label=(\roman*), labelsep=0.75em]
    \item Determine $u^0$ from the normalization condition $u^\mu u_\mu = -1$,
        namely solving the quadratic equation
        \begin{equation}
            g_{00} (u^0)^2 + 2g_{0i}u^i (u^0) + g_{ij}u^iu^j = -1 \, ,
        \end{equation}
        on each grid points. If a root $u^0$ does not exist, velocity is floored
        to $u^i=0$. Such failure only occurs in the region very close to the
        horizon ($r \lesssim 3r_g$), and the floored regions plunge into the BH
        as soon as a simulation begins, having negligible impact on the results.
    \item Add the BH recoil to the coordinate 3-velocity $u^i/u^0$ as\footnote{A
        more rigorous approach is applying the global Lorentz boost
        \cite[e.g.][]{Megevand:2009yx,Zanotti:2010xs} on the CBD matter profile
        into the rest frame of the recoiled BH. However, the relative error from
        our approximate treatment is in the order of $(W - 1)$, where $W =
        (1-v_r^2)^{-1/2}$ is the Lorentz factor corresponding to the recoil
        speed $v_r$. This amounts to 0.5\% for the largest recoil speed
        $v_r=0.10$ considered in this work and 0.1\% for $v_r=0.05$ mainly
        discussed in the paper, which we consider negligible regarding our main
        purpose.}
        \begin{equation}
            \left(\frac{u^i}{u^0}\right) \leftarrow \left(\frac{u^i}{u^0} - v_r^i\right) \, .
        \end{equation}
        The minus sign accounts for the fact that, in the rest frame of the BH,
        the CBD matter profile picks up a velocity with an opposite sign.
    \item Using the values of $u^i/u^0$ from the previous step, compute $u^{i'}$
        with
        \begin{equation}
            v^i \equiv \sqrt{-g^{00}} \left( u^i / u^0 + \beta^i \right) \, ,
        \end{equation}
        \begin{equation}
            W = (1-g_{ij}v^iv^j)^{-1/2} \, ,
        \end{equation}
        then $u^{i'} = Wv^i$.
    \end{enumerate}
\item Magnetic fields are imported $(\hat{\mathcal{B}}^i)_{\rm GR} \leftarrow
    (\hat{B}^i)_{\rm Newt}$ without any modification. Since our simulations are
    performed in the Cartesian Kerr-Schild coordinates, in which $\sqrt{-g} =
    1$, the Gauss constraint of the magnetic field $\partial_i (\sqrt{-g}
    \mathcal{B}^i_{\rm GR}) = 0$ simply reduces to $\partial_i
    \mathcal{B}^i_{\rm GR} = 0$. Since $(B^i)_{\rm Newt}$ from the imported
    numerical data readily satisfies $\partial_i B^i_{\rm Newt} = 0$ on its
    discrete grid, the divergence-free condition is seamlessly preserved,
    provided that the rescaling of spatial coordinates is uniform in all spatial
    directions.
\end{itemize}

\begin{figure*}
    \includegraphics[width=\linewidth]{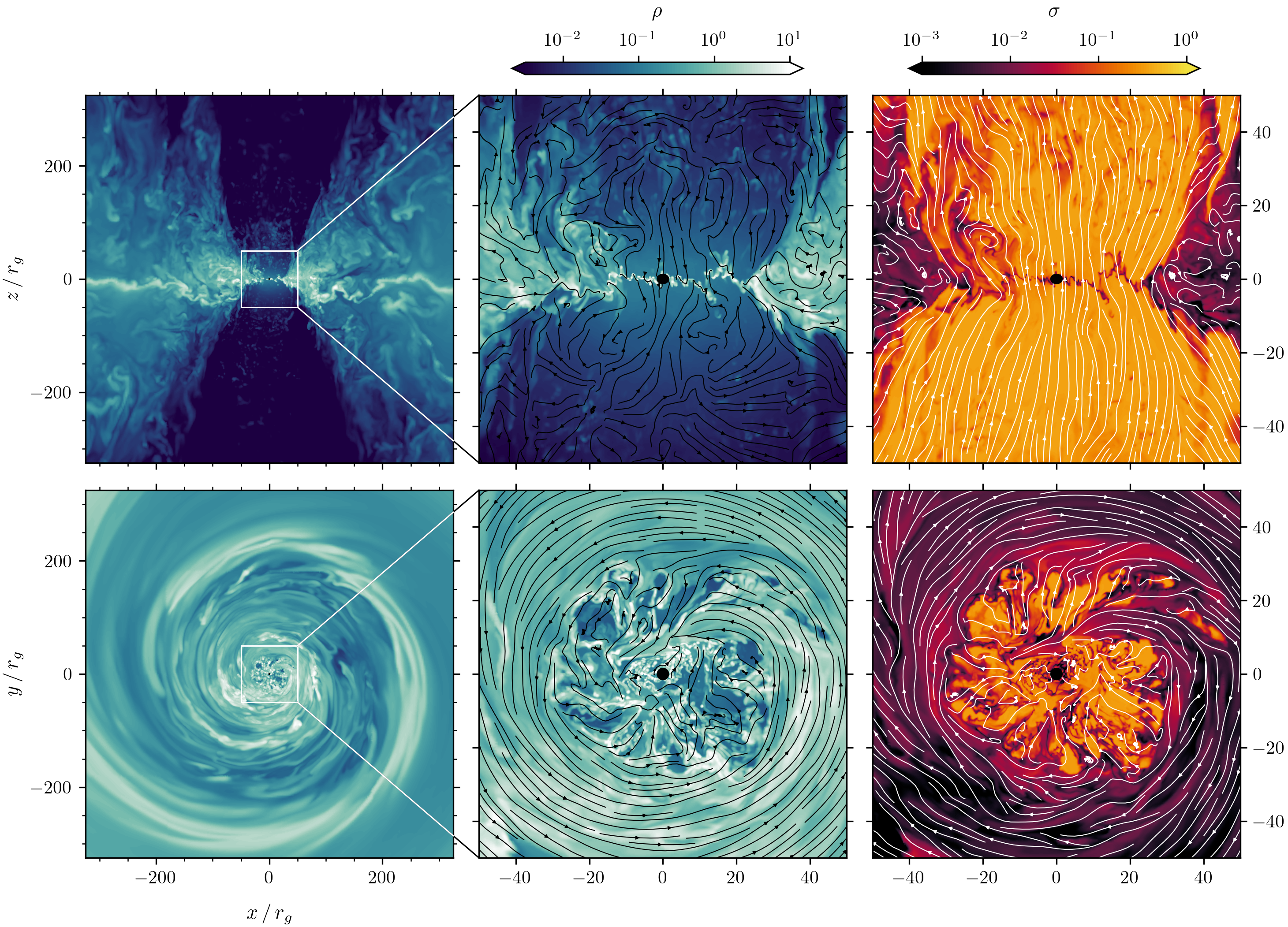}
    \caption{Initial data mapped from the Newtonian MHD simulation of
        \citetalias{Most:2024onq}: the rest mass density $\rho$ on the vertical
        plane ($xz$, first row) and midplane ($xy$, second row) over the global
        CBD structure (first column), and the region close to the BH (second
        column), along with the relativistic magnetization $\sigma$ (third
        column). In-plane velocity streamlines (second column, black solid
        lines) and magnetic field lines (third column, white solid lines) are
        overlayed. Shown here is an example initial data with no BH recoil
        added; see Sec.~\ref{sec:mapping} for how the BH recoil is added onto
        the matter profile.}
    \label{fig:initial data}
\end{figure*}

Fig.~\ref{fig:initial data} shows the initial data with $v_r = 0$ after the
mapping is completed. The degree of magnetization in \citetalias{Most:2024onq}
was capped in terms of the plasma beta parameter $\beta = 2p_{\rm gas} / B^2 >
10^{-3}$. After rescaling variables into the GRMHD units, we get the
relativistic magnetization in the CBD cavity of about $\sigma \approx 0.3$.

\section{Time Evolution}

Time integration of ideal GRMHD equations is performed using a second order
Runge-Kutta stepper, with the spatial discretization using piecewise parabolic
reconstruction \cite{Colella:1982ee} and the HLLE Riemann fluxes
\cite{Harten1983,Einfeldt1991}. Magnetic fields are evolved with the constrained
transport algorithm \cite{Evans:1988qd,Gardiner:2007nc}. We limit the maximum
Lorentz factor of the fluid to $W_{\rm max} = 20$, and cap the maximum
magnetization $\sigma_{\rm max} = 50$ using the drift flooring algorithm
\cite{Ressler:2016pmh}. No inflow boundary condition is imposed at the domain
boundaries.

\section{Mass accretion rate and magnetic fluxes}

\begin{figure*}
    \vspace{1em}
    \center
    \includegraphics[width=\linewidth]{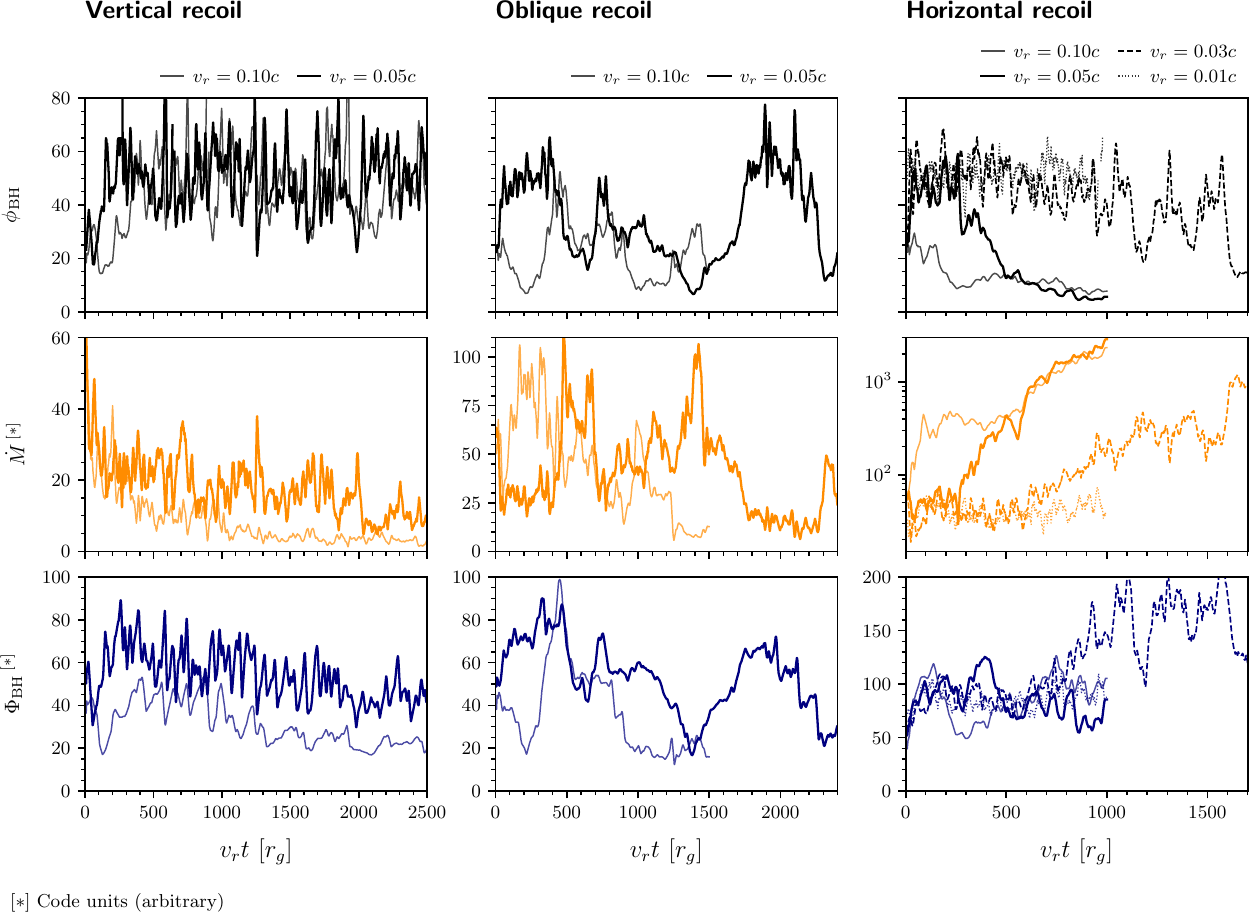}
    \caption{Time evolution of the dimensionless MAD parameter $\phi_{\rm BH}$,
    mass accretion rate $\dot{M}$, and horizon magnetic fluxes $\Phi_{\rm BH}$
    from all simulations.
    The horizontal axes (simulation time) are normalized in the units of $v_r
    t$, the coordinate distance the recoiled BH has traveled.
    A zeroth order Savitzky-Golay filter is applied over the interval
    $\Delta(v_r t) = 15r_g$ for smoothing the time series data shown here.}
    \label{fig:history}
\end{figure*}

In Fig.~\ref{fig:history}, for all simulations, we plot the rest mass accretion
rate
\begin{equation} \label{integral-mdot}
    \dot{M} = \oint (-\rho u^r) \sqrt{-g} \, d\theta d\phi \, ,
\end{equation}
total magnetic fluxes threading the BH
\begin{equation} \label{integral-Phi}
    \Phi_\text{BH} = \frac{1}{2} \oint |\dual{F}^{i0}| \sqrt{-g} \, d\theta d\phi \, ,
\end{equation}
and the dimensionless MAD parameter
\begin{equation} \label{phi-mad}
    \phi_{\rm BH} \equiv
    \left(\frac{4\pi \Phi_{\rm BH}^2}{\dot{M} r_g^2 c}\right)^{1/2} \, ,
\end{equation}
where $\rho$ is the rest mass density, $u^r$ is the radial component of the
four-velocity, and $g$ is the determinant of the spacetime metric. $\phi_{\rm
BH} \gtrsim 50$ characterizes the magnetically arrested disk (MAD) state
\cite{Narayan:2003by,Igumenshchev:2007bh} accompanying relativistic jets
launched from the BH \cite{Tchekhovskoy:2011zx}. The mass accretion rate
$\dot{M}$ is extracted at $r=3r_g$ to reduce contamination from numerical
flooring, and $\Phi_{\rm BH}$ is computed on the outer horizon.

\section{Total bound mass and accretion lifetime}

\begin{figure}
\center
\includegraphics[width=0.55\textwidth]{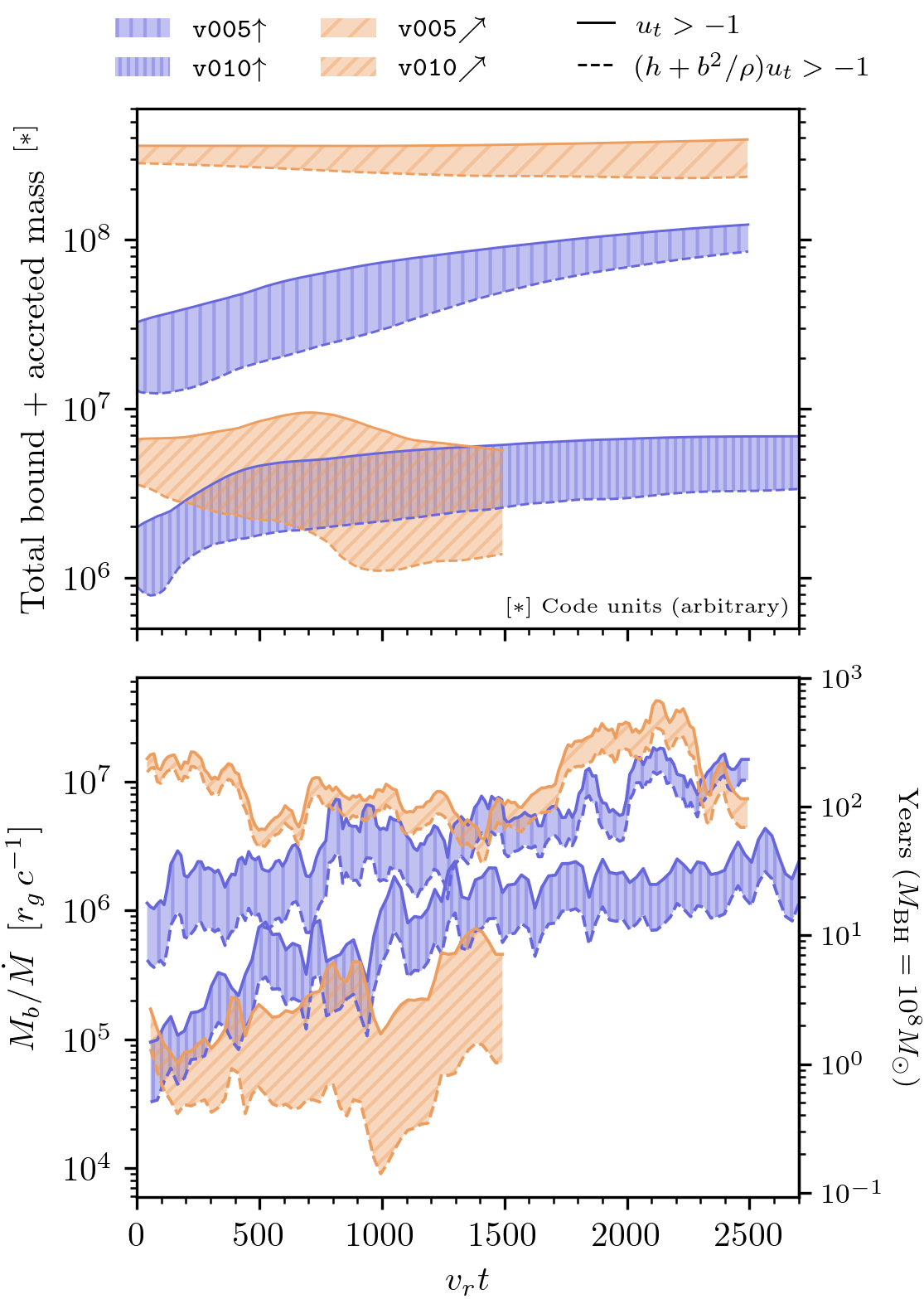}
\caption{Diagnostic quantities on the mass in the vertical and oblique recoil
    models. (Top) Total mass of the bound and accreted material. (Bottom) The
    (instantaneous measure of) accretion lifetime $M_b / \dot{M}$. Lower and
    upper boundaries of each shaded regions used the boundness criteria
    \eqref{eq:bernoulli-particle} and \eqref{eq:bernoulli-mhd} for computing the
    total bound mass $M_b(t)$, respectively.}
\label{fig:accretion-lifetime}
\end{figure}

We compute the total bound mass $M_b(t)$ at a given simulation time $t$ based on
the geodesic criterion
\begin{equation} \label{eq:bernoulli-particle}
    -u_t < 1 \, ,
\end{equation}
and the MHD Bernoulli criterion \cite{Penna:2013zga}
\begin{equation} \label{eq:bernoulli-mhd}
    - \left(h + \frac{b^2}{\rho}\right) u_t < 1 \, ,
\end{equation}
where $\rho$ is the rest mass density, $h$ is the specific enthalpy, and $b^2$
is the comoving magnetic energy density. The geodesic criterion
\eqref{eq:bernoulli-particle} ignores the internal energy of the fluid and
generally overestimates the bound mass, where the Bernoulli crietrion
\eqref{eq:bernoulli-mhd} can underestimate \cite[see,
e.g.][]{Kastaun:2014fna,Foucart:2021ikp}. Further, the Bernoulli criterion
\eqref{eq:bernoulli-mhd} is only valid for a stationary flow, hence it only
gives a crude estimate given the highly dynamic state of the gas flow in our
simulations. In this regard, we present the bound mass from both criteria in
order to gauge the uncertainty and place brackets.

The top panel of Fig.~\ref{fig:accretion-lifetime} shows sum of the total bound
mass $M_b(t)$ and the total rest mass accreted $M_{\rm acc}(t) \equiv \int_0^t
\dot{M}(t') dt'$ in the vertical and oblique recoil models, essentially showing
the total mass of the material that has been ejected out from the CBD by the
recoiled BH. In the models with $v_r=0.10$ (\texttt{v010$\uparrow$},
\texttt{v010$\nearrow$}), the accreted mass is somewhat comparable to the total
bound mass in a later time (up to a few of 10\%), where in the models with
$v_r=0.05$ (\texttt{v005$\uparrow$}, \texttt{v005$\nearrow$}), its fraction is
negligible ($\lesssim 1\%$). The bound mass in \texttt{v010$\uparrow$} and
\texttt{v010$\nearrow$} models converge and show turnover within the simulation
time. Slow but indefinite increases of the total bound mass in the
\texttt{v005$\uparrow$} and \texttt{v005$\nearrow$} models are owing to the
weakly bound ejecta.

The bottom panel of Fig.~\ref{fig:accretion-lifetime} is plotting
$M_b(t)/\dot{M}(t)$, which is an instantaneous measure of the BH accretion
lifetime. Short-time fluctuations are coming from the variability of the mass
accretion rate in the MAD state.

\end{document}